\documentstyle[12pt]{article}
\newcommand{\comma}{\;\;\;\; ,}
\newcommand{\period}{\;\;\;\; .}
\newcommand{\eq}{\; = \;}
\newcommand{\sep}{\;\;\; , \;\;\;}
\newcommand{\be}{\begin{equation}}
\newcommand{\bd}{\begin{displaymath}}
\newcommand{\ee}{\end{equation}}
\newcommand{\ed}{\end{displaymath}}
\newcommand{\ba}{\begin{eqnarray}}
\newcommand{\ea}{\end{eqnarray}}

\title{ A direct proof of Kim's identities}

\author{ R.J. Baxter$^1$ \\
{\protect \small Theoretical Physics, I.A.S. and School of Mathematical
Sciences}\\
{\protect \small  The Australian National University,
 Canberra, A.C.T. 0200, Australia  }
}
\date{}

\begin{document}

\maketitle

\abstract{
As a by-product of a finite-size Bethe ansatz calculation in statistical mechanics,
Doochul Kim has established, by an indirect route, three mathematical identities rather 
similar to the conjugate
modulus relations satisfied by the elliptic theta constants. However, they contain
factors like $1 - q^{\sqrt{n}}$ and $1 - q^{n^2}$, instead of $1 - q^n$.  We show here that there
is a fourth relation that naturally completes the set, in much the same way that there are four
relations for the four elliptic theta functions. We derive all of them directly by proving 
and using a specialization of Weierstrass' factorization theorem in complex variable theory.}
\vspace{6cm}
{\protect \footnotetext[1]{This work was partially supported
by the Asia-Pacific Center for Theoretical Physics, Seoul, Korea\\
To appear in J. Phys. A (1998).}}
\pagebreak

\subsection*{Introduction}

 Kim (1996) has obtained the leading finite-size corrections to the spectra of the 
asymmetric XXZ chain and the related six-vertex model, near the antiferromagnetic 
phase boundary at zero vertical field. He also performs the calculation at zero 
horizontal field. These results are related by a $90^{\circ}$
rotation, which yields the following three identities, true for all real positive
$\tau$:
\ba \label{ident1}
 \prod_{n = -\infty} ^{\infty} 
\left[1 + p^{(2 n - \! 1)^2} \right] & \eq &    ( (q \overline{q} ) ^{-c} \prod_{n = 1} ^{\infty}
\left(1 + q^{\sqrt{2 n  - \!1}} \, \right) ^2 \; \left(1 + \overline{q}^{\sqrt{2 n  - \!1}} \, \right) ^2 
 \\
\label{ident2}
 \prod_{n = -\infty} ^{\infty}
\left(1 + p^{4 n^2} \right) & \eq &   (q \overline{q} ) ^{-c} \prod_{n = 1} ^{\infty}
\left(1 - q^{\sqrt{2 n - \! 1}} \, \right) ^2 \; \left(1 - \overline{q}^{\sqrt{2 n - \! 1}} \, \right) ^2 \\
\label{ident3}
   \prod_{n = -\infty} ^{\infty}
\left[ 1 - p^{(2 n - \!1)^2} \right]   & \eq &   4 \, (q \overline{q} ) ^{d} \prod_{n = 1} ^{\infty}
\left(1 + q^{\sqrt{2 n}} \, \right) ^2 \; \left(1 + \overline{q}^{\sqrt{2 n}} \, \right) ^2  \ea

Here
\bd
 c  = (\surd 2 - 1) \; \zeta(3/2)/(4 \pi  )  = 0.08610929... \comma \ed
\bd  d   = \zeta(3/2)/(2 \pi \surd 2 )  = 0.29399552... \sep 
 p  = e^{ - \pi /\tau } \comma \ed
\bd
q  \eq  e^{-\pi \sqrt{i \tau} } \sep 
\overline{q} \eq e^{-\pi \sqrt{-i \tau}} \period \ed
All square roots herein are chosen to be in the right half plane.
By $q^{\lambda}$,  $\overline{q}^{\lambda}$ we mean 
$ e^{-\pi \lambda \sqrt{i \tau} }$, $ e^{-\pi \lambda \sqrt{-i \tau} }$,
respectively.

These identities (\ref{ident1}) -- (\ref{ident3}) are reminiscent of the conjugate modulus
identities of the  elliptic theta constants $H_1(0), \Theta(0),
\Theta_1(0)$ [section 21.51  
of Whittaker and Watson (1950),
p 75 of  Courant and Hilbert (1953), eqn 15.7.2 of Baxter(1982)].
However, in those identities $q$ ( or rather $e^{-\pi \tau}$) and $p$ are raised to a power proportional to $n$,
rather than $n^2$ or $\sqrt{n}$. We should like to have a direct proof of
(\ref{ident1}) -- (\ref{ident3}). We obtain one here by using the Poisson transformation and complex
variable theory to establish the general result (\ref{gen}). We specialize this to 
(\ref{allfour}), from which we obtain a fourth identity:
\be \label{ident4}
 \prod_{n=1}^{\infty} \left( 1 - p^{4 n^2} \right) ^2  \eq
 \pi \tau   \; (q \overline{q})^{d } \prod_{n=1}^{\infty}
 \left(1 - q^{\sqrt{2 n}} \, \right) ^2  \left( 1 - \overline{q}^{\sqrt{2 n}} \, \right) ^2
 \period \ee
This is analogous to the conjugate modulus identity for $H'(0)$.

We can write the products in (\ref{ident1}) -- (\ref{ident3}) in terms of the type of products
occurring in (\ref{ident4}) by using the elementary identities
$1 + x = (1-x^2)/(1-x)$ and $\prod_n \, f(2 n-1) = \prod_n \, [f(n)/f(2n)]$.
Write identity $(j)$, for a given value of $\tau$, as $(j,\tau)$. 
Then in this way we find that 
$(2,\tau)$ can be obtained from the ratio $(4,\tau /2):\!(4,\tau )$.
Also, $(3,\tau)$ follows from  $(4, 4\tau )\! : \! (4,\tau )$. Finally, 
$(1,\tau)$ can be obtained from the ratio $(2, 4\tau)\! : \!(2,\tau)$ , or
alternatively from $(3,\tau /2) :\! (3,\tau )$. Thus (\ref{ident4}) implies
 (\ref{ident1}) --  (\ref{ident3}).

\subsection*{Proof of identity (\ref{ident4})}

We begin by proving a general theorem. Let $F(z)$ be a meromorphic function 
of a complex variable such that

(i)  $\log F(z)$ is analytic on the real axis and $\log F(x)$ is Fourier 
 analyzable;

(ii) the integral $\int_{-\infty}^{\infty} e^{i k x} F'(x)/F(x)$ can be closed round
the upper half plane for ${\rm Re}\, k > 0$, round the lower half plane for 
 ${\rm Re}\, k < 0$ (i.e. there exists a discrete sequence of ever increasing 
appropriate arcs such that the integral over the arc tends to zero);

(iii) The zeros and poles of $F(z)$ in the UHP are at $u_1, u_2, \ldots$, and $m_r$
is the  multiplicity of the zero at $u_r$ (regarding poles as zeros with 
negative multiplicity: thus a pole of order $j$ has multiplicity $-j$). Similarly,
the zeros and poles in the LHP are at $v_1, v_2, \ldots$, with multiplicities
$n_1, n_2, \ldots$.

Then, for all real positive $x$ and $\delta$,
\begin{eqnarray} \label{gen}
\prod_{n=-\infty}^{\infty} F(x + n \delta ) \; \; &  = & \; \; 
\exp \left[ \delta^{-1} \int_{-\infty}^{\infty} \log F(t) \; dt \right] \; \;  
\prod_r \left[ 1 - e^{2 \pi i (u_r - x)/\delta } \right]^{m_r} \nonumber \\
& &
\times \prod_{r} \left[ 1 - e^{2 \pi i (x - v_{r})/\delta } \right]^{n_{r}} \comma \end{eqnarray}
the first product on the RHS being over the zeros and poles in the UHP, the 
second over those in the LHP.

This identity is a variant of Weierstrass' factorization theorem
(section 7.6 of 
Whittaker and Watson, 1950). 
One can readily verify
that both sides, considered as  functions of $x$, have the same zeros
and poles. 

\paragraph{Proof}
To prove this theorem, let $g(k)$ be the Fourier transform of $\log F(x)$:
\be \label{Fint}
g (k) = \int_{-\infty}^{\infty} \, e^{i k x} \, \log F(x) \, dx \ee
By ``Fourier analyzable'' in requirement (i) we mean that this integral
is absolutely convergent, for all real $k$.

Then from the Poisson transform (pp. 75 -- 77 of Courant and Hilbert, 1953; 
eqn. 15.8.1 of Baxter, 1982)
applied to the function $\log F(x + a)$, where $a$ is an arbitrary real parameter,
\be \label{poiss}
 \sum_{n=-\infty}^{\infty} \; \log F(a + n \delta ) \eq \delta^{-1}
\sum_{n=-\infty}^{\infty} \; e^{-2 \pi i n a/\delta } \, g( 2 \pi n /\delta )  \ee

Integrating (\ref{Fint}) by parts, noting that $\log F(x)$ necessarily
tends to zero as $x \rightarrow \pm \infty$,
\be
g (k) \eq (i/k) \int_{-\infty}^{\infty} \, e^{i k x} \, F'(x)/F(x) \, dx \period \ee
If $k > 0$, it follows from requirement (ii) that
\be g(k) \eq - \frac{2 \pi }{ k} \; \sum_r \; m_r \, e^{i k u_r} \comma \ee
while for $k < 0$
\be g(k) \eq \frac{2 \pi }{ k} \; \sum_r \; n_r \, e^{i k v_r} \comma \ee
the sums being over the zeros $u_r$ and $v_r$, respectively.

Substituting these expressions into the RHS of (\ref{poiss}) and interchanging
the order of the summations, we obtain
\be
 \sum_{n=-\infty}^{\infty} \; \log F(a + n \delta ) \eq \frac{g(0)}{\delta} + 
\sum_r m_r \log [ 1 - e^{2 \pi i (u_r - a)/\delta } ] +
\sum_r n_r \log [ 1 - e^{ 2 \pi i (a -  v_r)/\delta } ] \period \ee
Exponentiating and replacing $a$ by $x$, we obtain the desired result (\ref{gen}).

\paragraph{Corollaries}

Let $y$ be real, in the range $-1 < y < 1$. Define
\be
s  \eq e^{i \pi y} \sep w = e^{i \pi x}  \comma\ee
\be
F(z) = 1 + s \, e^{ - \pi z^2/\tau} \period \ee

Regard $y$ and $\tau$ as constants:
then $F(z)$ is entire, with only simple zeros, and
satisfies conditions (i) - (iii). Set $\delta = 2$ and use the above definitions of $p$, $q$,
$\overline{q}$. Then (\ref{gen})  becomes,
 for all real $x$,
\ba \label{allfour}
\prod_{n = -\infty}^{\infty} \left[ 1 + s \, 
p^{(2 n + x)^2} \right]
& \eq  & (q \overline{q})^{- h(s)} \prod_{n = 1}^{\infty}  \left\{ \left[
1 - w \, q^{ \sqrt{2 n -1 - y} } \right] 
\left[ 1 - w^{-1} q^{ \sqrt{2 n -1 - y} } \right] \right. \nonumber \\
& & \times  \left. \left[
1 - w \, \overline{q}^{ \sqrt{2 n - 1 + y} } \right] 
\left[ 1 - w^{-1} \overline{q}^{ \sqrt{2 n  - 1 + y} } \right] \right\}
\comma \ea
where
\be h ( s ) \eq    (2 \pi)^{- 3/2} \int_{-\infty}^{\infty} 
\log \left( 1 + s \, e^{- t^2 } \right)
\, dt  \eq    \sum_{r=1}^{\infty} \frac {(-1)^{r-1} \,s^r}{\pi \, (2 r)^{3/2}} 
\period \ee
Thus $h(-1) = - d$, $h(1) = c$.

Setting $x = 0$ and letting $y \rightarrow 1$  , we obtain the identity 
(\ref{ident4}). (The $n=0$ factor on the LHS, 
and the first two $n=1$ factors on the RHS, vanish in this limit: we have to evaluate
them to leading non-zero order and then take their ratio.) 

As we have indicated above, this is sufficient to establish Kim's three identities 
(\ref{ident1}) -- (\ref{ident3}). However, we also note that (\ref{allfour}) contains all of 
(\ref{ident1}) -- (\ref{ident4}) as special cases:
(\ref{ident1}) can be obtained  by setting $x = -1$ and $y = 0$; (\ref{ident2})
by setting $x = y = 0$; and (\ref{ident3})
by setting $x = -1$ and $y = 1$.

The relation (\ref{allfour}) plays a similar role to the conjugate modulus relation satisfied 
by the elliptic theta functions for arbitrary values of their argument. It is  interesting 
to speculate whether the products therein,
considered as functions of $x$ or $y$, 
have any other properties resembling elliptic functions. 
For instance, are there algebraic relations corresponding to
${\rm sn}^2 u + {\rm cn}^2 u = 1$ or $k^2 + k'^2 = 1$? 

\subsection*{Acknowledgement}

The author thanks  Doochul Kim for drawing his attention to these 
intriguing identities. 

\subsection*{Note added in proof}

  The author is indebted to George E. Andrews for pointing out that 
equation (4) is stated (without reference or derivation)  by G.H. Hardy 
and S. Ramanujan in section 7.3 of their paper on Asymptotic Formulae 
in Combinatorial Analysis, Proc. London Math. Soc. (2) 17 (1918) 75 - 115.
This is reprinted in volume 1 of G.H. Hardy, Collected Papers, 
Clarendon Press, Oxford (1966), p 335.

\subsection*{References}

\noindent Baxter R J 1982 {\it Exactly Solved Models in Statistical Mechanics}
(London: Academic)

\noindent Courant R  and Hilbert D 1953 {\it Methods of Mathematical Physics}
Vol 1 (New York: Interscience)

\noindent Kim D 1996 {\it J. Phys. A: Math. Gen.} {\bf 30} 3817 -- 3836

\noindent Whittaker E T and Watson G N 1950 {\it A Course of Modern Analysis} 
(Cambridge University Press)

\end{document}